# Autotuning T-PaiNN: Enabling Data-Efficient GNN Interatomic Potential Development via Classical-to-Quantum Transfer Learning

Vivienne Pelletier[1], Vedant Bhat[2], Daniel J. Rivera[2], Steven A. Wilson[2], Christopher L. Muhich[1,2]*

[1] Materials Science & Engineering, School for the Engineering of Matter, Transport, & Energy, Arizona State University, 551 E. Tyler Mall, Tempe Arizona, 85287, USA
[2] Chemical Engineering, School for the Engineering of Matter, Transport, & Energy, Arizona State University, 551 E. Tyler Mall, Tempe Arizona, 85287, USA
* Corresponding Author

## Abstract

Machine-learned interatomic potentials (MLIPs), particularly graph neural network (GNN)-based models, offer a promising route to achieving near–density functional theory (DFT) accuracy at significantly reduced computational cost. However, their practical deployment is often limited by the large volumes of expensive quantum mechanical training data required. In this work, we introduce a transfer learning framework, Transfer-PaiNN (T-PaiNN), that substantially improves the data efficiency of GNN-MLIPs by leveraging inexpensive classical force field data. The approach consists of pretraining a PaiNN MLIP architecture on large-scale datasets generated from classical molecular simulations, followed by fine-tuning (dubbed autotuning) using a comparatively small DFT dataset. We demonstrate the effectiveness of autotuning T-PaiNN on both gas-phase molecular systems (QM9 dataset) and condensed-phase liquid water. Across all cases, T-PaiNN significantly outperforms models trained solely on DFT data, achieving order-of-magnitude reductions in mean absolute error while accelerating training convergence. For example, using the QM9 data set, error reductions of up to 25× are observed in low-data regimes, while liquid water simulations show improved predictions of energies, forces, and experimentally relevant properties such as density and diffusion. These gains arise from the model's ability to learn general features of the potential energy surface from extensive classical sampling, which are subsequently refined to quantum accuracy. Overall, this work establishes transfer learning from classical force fields as a practical and computationally efficient strategy for developing high-accuracy, data-efficient GNN interatomic potentials, enabling broader application of MLIPs to complex chemical systems.

## 1. Introduction

The development and use of machine learned interatomic potentials (MLIPs) is rapidly expanding due to their promise for overcoming the chemical accuracy/computational speed tradeoff that has defined computational atomistic modeling since its inception.[1] MLIPs are appealing surrogate models for acquiring near-quantum mechanical, e.g. density functional theory (DFT), accuracy at near-classical molecular mechanics costs.[2] MLIPs have been developed as both stand-alone models, for example in molecular dynamics simulations,[3] and as complimentary DFT accelerators, for example in geometry optimization and transition state searches.[4][5] Among the broad array of candidate machine learning architectures used for MLIPs, two predominant categories have emerged: 1) Kernel-regression based (Kernel-MLIP), and 2) graph neural network (GNN),[6] due to their ability to broadly reproduce DFT level accuracy, transferability, and size agnosticism.

Kernel-MLIPs such as Gaussian Approximation Potentials[2] and their derivatives, e.g. VASP's MLFFs,[7] are appealing DFT surrogates because of their strong performance in the low training data regime. Kernel-MLIPs predict system energies by calculating the similarity between the atomic structure of interest and all structures in the training set via a similarity function, the kernel. Energy and force predictions of new structures are constructed from a learned linear combination of similarities to the training data. Forces and stresses are calculated from the positional and cell lattice vector analytical derivatives of this procedure. The most popular Kernel-MLIP methods employ Bayesian inference to provide an inbuilt self-assessment of their own uncertainty, thus enabling both retrospective assessment of calculation results and active learning "on-the-fly" launching of DFT calculations.

Kernel-MLIP memory demands scale poorly, typically as $\mathcal{O}(N^2)$, with training set size.[8] Although this scaling may remain manageable for narrowly targeted investigations, it becomes prohibitive when the objective is to treat many dissimilar candidate molecules, solid materials, surface terminations, etc. While methods for reducing the memory demands of Kernel-MLIPs exist, such as sparse inducing points and Nyström approximations, these approaches only improve the unfavorable scaling from $\mathcal{O}(N^2)$ to $\mathcal{O}(NM^2)$, where $M$ is related to the sparsity of the approximation.[9] In practice, to maintain the necessary accuracy for MLIP applications, $M$ must be chosen so large that the improvement in memory scaling is ultimately modest.[8] Thus, alternative MLIP architectures must be used if broad applicability is desired.

GNN-MLIPs predict molecular system energies and properties using a series of neural networks, and/or their derivatives (i.e. convolutions, perceptron, etc.).[10] Each atom is first assigned a descriptor vector, typically initialized using its element type, then a GNN performs iterative updates of these embeddings based on interactions with neighboring atoms in the atom's chemical environment. Lastly, the atomic descriptors feed a NN that predicts each atom's individual contribution to the system energy (or other desired property). GNNs encompass a range of architectures, such as SchNet,[10] Polarizable Atom Interaction Neural Network (PaiNN),[11]

NequIP,[12] and GemNet,[13] and rank as the best performing models to-date on test sets, such as the Open Catalyst Project and Direct Air Capture dashboards.[14][15]

GNN-MLIPs do not suffer from quadratic memory scaling with respect to the number of training samples but require significantly larger training sets to fit their large number of parameters (tens of thousands to hundreds of thousands). GNN model size is fixed by the GNN architecture, i.e. the number and sizes of GNN embeddings, convolutional, and predictive layers, etc. The sizes of the matrices representing various convolutions and activations remain constant across systems as they describe individual elements and atoms. Thus, GNNs are extendable to any size system and have a constant execution cost with respect to target size regardless of training set size. However, the modest DFT datasets sizes that are often sufficient for constructing a high-performing GAP MLIP are generally insufficient to train a GNN MLIP of comparable accuracy and robustness. Therefore, the central difficulty in construction of MLIPs is that the model class with the most favorable scaling to large problems (GNN-MLIPs) is also the one whose data requirements most strongly conflict with the cost of DFT data generation. Thus, improvements to MLIPs for large scale diverse simulations require an improvement in training data collection efficiency of GNN-MLIPs.

This contribution aims to bridge the knowledge gap of how to achieve high quality GNN training with minimal DFT training data using transfer learning. Transfer learning is a machine learning technique where a model trained on one task is reused as the initial state for training on another related task.[16] Neural networks have been demonstrated to benefit from transfer learning in many diverse tasks, such as image classification,[17] machine translation,[18] and 3D pose estimation from 2D images,[19] as just a few examples. To-date, use of transfer learning in atomistic systems has been restricted to using models pre-trained on large scale generalized DFT datasets and then refined for systems of interest, e.g. Δ-ML.[20] This transfer learning application, therefore, still relies on the collection of sufficient DFT data to fully span the configurational space of the system of interest.

This work demonstrates a novel transfer learning paradigm for achieving high quality GNN-MLIPs with minimal DFT training data by using classical force fields, which we call Transfer-PaiNN, or T-PaiNN, because we employ the PaiNN ML architecture. Specifically, we improve the DFT data efficiency of GNN-MLIPs by pretraining a PaiNN model on a large dataset (>100 times the DFT data set) generated from classical force field simulation of the system of interest and then tuning the model weights using the small DFT data set. Classical force fields provide a computationally inexpensive approximation of the DFT potential energy surface (PES) and are directly usable in molecular dynamics simulations for generating the large datasets necessary to train robust GNN-MLIP. Because classical force field simulations are effectively free in terms of computational time compared to DFT simulations, the computational cost of this extra step is minimal. We find that our T-PaiNN methodology not only significantly decreases the DFT data required to train high performance MLIPs, but *it also improves* the generality of the MLIPs performance.

## 2. Methods

This section details the fundamental concept used in T-PaiNN methodology, the underlying graphical network architecture, the exemplar systems tested, and the computational data collection methods, including both the classical and quantum calculation methods.

### 2.1 Transfer learning for Graph Neural Network Prediction of Atomistic Systems.

The central hypothesis underlying the T-PaiNN workflow is that there exists a strong correlation between atomic geometries and the associated energy/force data set calculated at both the classical and quantum levels because they are both methods of modeling the behavior of atomic systems, albeit with varying levels of accuracy and cost. Therefore, the information embedded into MLIP models trained on classical force fields are directly translatable to quantum mechanical tasks. This premise can be placed into a statistical framing as follows. Given two data sets, a classical force field data set $\mathcal{D} = \{(x_i, y_i)\}_i^N$, where $x_i \in \mathcal{X}$ is a feature vector representing atomic structures drawn from sampling distribution $x \sim p(\mathcal{X})$ with $y_i \in \mathcal{Y}$ as the corresponding energy/force label from the target space $\mathcal{Y}$ calculated by classical force fields, and a quantum data set, $\mathcal{D}'$, consisting of $x_i' \sim p'(\mathcal{X})$ labeled as $y_i' \in \mathcal{Y}'$ calculated at the quantum level, if a model can map $f: \mathcal{X} \to \mathcal{Y}$, only minimal information about $\mathcal{D}'$ is necessary to retrain $f$ to mapping $f': \mathcal{X}' \to \mathcal{Y}'$.

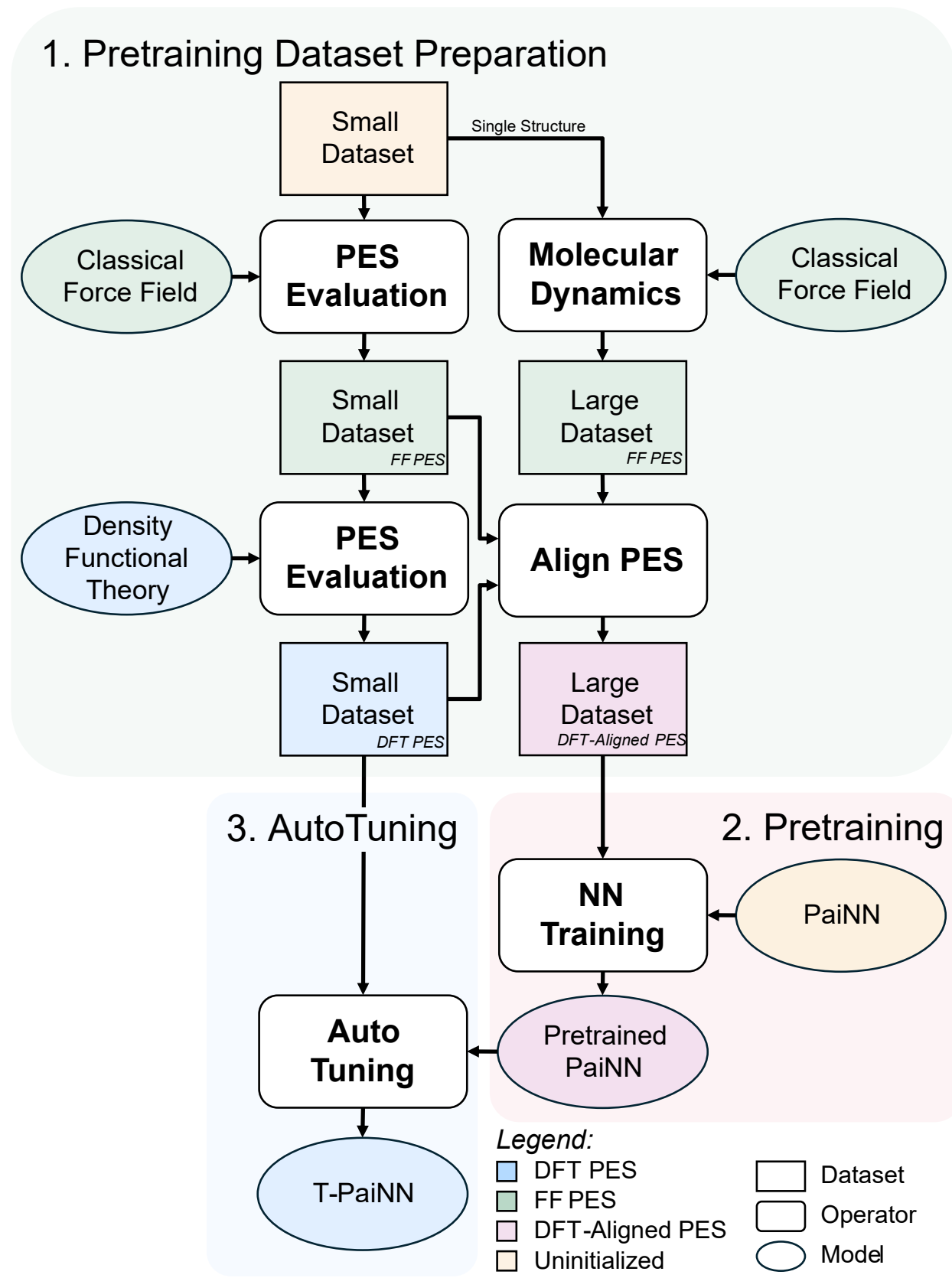

**Figure 1**: Schematic of the T-PaiNN procedure, with shapes representing each object's type and colors representing the quality of the PES for that object.

The application of the T-PaiNN workflow progresses in three main stages, as shown schematically in Figure 1: 1) the generation of a classical force field dataset on the system of interest, 2) pretraining the GNN by the classical forcefield data set, then 3) the transfer learning finetuning of the GNN using DFT results. We refer to this last stage as Autotuning T-PaiNN because it tunes the T-PaiNN model weights based on the DFT information used to align the classical and DFT information, and thus this information is automatically provided.

The classical mechanics derived pretraining dataset represents a more complete sampling of an approximate PES than is possible using DFT, while the DFT dataset represents a sparse sampling of a more accurate PES. The pretraining stage takes advantage of this broader sampling by placing the model's weights into a region of parameter space which is consistent with a smooth and physically plausible PES. With the weights initialized in this state, autotuning T-PaiNN on the DFT data adjusts the weights to be consistent with the DFT PES without leaving this smooth-PES region of the space. Without pre-training and autotuning, DFT-only fitting must discover both the organization of configuration space and the mapping from the embedding to the DFT PES from a much more limited set of data. On subsequent evaluations where extrapolations are unavoidable, such as when performing molecular dynamics simulations, the T-PaiNN models are hypothesized to remain more robust because they were trained on a far greater number of atomic configurations than equivalent models only trained on DFT data (i.e. DFT-only PaiNN models).

We note two important points necessary to maximize the performance of T-PaiNN. Firstly, after the classical dataset has been generated, the force field and DFT potential energy surfaces (which are defined only to an additive constant) must be aligned using paired force field and DFT evaluations on a shared set of structures. This alignment is done without violating any derived quantities (forces, stresses) via learned per-atom energy offsets for each element type derived from the residuals between force-field and DFT energies, and thus is essentially computationally free. Second, a classical force field which is appropriate for the system of interest should be selected to generate the pretraining data. It is preferable to utilize an existing force field that is applicable to the system of interest whenever possible, since force fields that have already been validated for molecular dynamic simulations of the systems of interest are more likely to remain numerically robust over the trajectories needed for dataset generation and are likely to better represent the PES. If such a classical force field does not exist, however, fitting standard classical force field functional forms to the available DFT data is also possible. Even a relatively simple fitted force field may still provide an informative approximation to the DFT potential energy surface due to the physical structure imposed by the parameterized functional forms.

### 2.2 Choice of GNN MLIP Architecture: Polarizable Atom Interaction Neural Network

The Polarizable Atom Interaction Neural Network (PaiNN) implementation of GNN-MLIP architecture is used to predict energies, forces and stresses from nuclear positions and is the chosen architecture of this work. However, we note that the transfer approach is agnostic to architecture and is generally applicable to all GNN-MLIPs. PaiNN implements a Deep Sets approach;[21] specifically, it embeds each atom into a latent representation space, performs interaction operations between atoms within this space according to the graph connectivity of the atoms, and performs final inference by collapsing these atomic embeddings into the quantity of interest, in our case, system energy. Interactions between atoms in the latent space are executed using a message passing GNN. Massage passing GNN's provide enough complexity to perform favorably on benchmark datasets[15] while remaining modest in parameter size, thus making its training more feasible.

In this work, we construct the PaiNN architecture with the hyperparameters reported in Table 1. We note that some models available online exceed these parameter counts by a factor of 10,; however, because the purpose of this contribution is focused on gains enabled by information transfer rather than maximizing performance, we use the more modest PaiNN size to avoid the additional computational cost of larger model training.

**Table 1**: PaiNN Hyperparameters for the 3 model sizes used in this study.

| | Small Model | Medium Model | Large Model |
|---|---|---|---|
| Atomic Embedding Dimension | 32 | 64 | 128 |
| Interaction Updates | 3 | 4 | 5 |
| Radial Basis Functions | 16 | 32 | 32 |
| Parameter Count | 36,300 | 197,000 | 534,000 |

### 2.3 T-PaiNN Test Systems

Two test systems were selected to demonstrate the performance of T-PaiNN method: 1) gas phase molecular systems taken from the QM9 database,[22] and 2) liquid phase bulk water calculated in this work. These systems were chosen to represent major classes of chemical systems examined in atomistic modeling while maintaining computational tractability by using the extant QM9 dataset and limiting the number of atom types in the other system. Each of these systems is discussed further below, including the computational parameters used for the models.

#### *2.3.1 Gas Phase Molecular Systems – The QM9 Data Set*

Gas phase molecular systems were extracted from the QM9 molecular energy database.[22] The QM9 database compiles ~134,000 molecular structures containing the elements H, C, N, O, and F and their energies as calculated by an array of different accuracy quantum chemical methods. We chose to use the QM9 DFT data calculated at the B3LYP exchange correlation functional level of DFT with a 6-31G(2df,p) basis set.[23][24] The universal force field (UFF)[25] classical force field was used to calculate the energies of these molecules in the LAMMPS program and was used for pretraining the T-PaiNN model.[26] Given the large number of structures, only single point energies were collected from the QM9 tabulated geometries; no geometry optimizations or molecular dynamics runs were used to gather a large data set. The accuracy measure of the various models on the QM9 dataset was selected to be the molecular energy of the system.

#### *2.3.2 Liquid Phase System – Bulk Water Data Set*

Bulk liquid phase water was selected as a representative for liquid systems because of its ubiquity in chemistry and biology, and its complex behavior relative to its small size, which arises from the complex combined ionic and covalent character of its hydrogen bonding network. The TIP3P water force field served as the classical forcefield for pretraining T-PaiNN.[27] While this force field is typically used with a rigid constraint placed on the water molecules, which prohibits O-H vibrations and H-O-H angle extensions, a flexible version of this force field was chosen to better align the results with the DFT target where the angle and bond lengths are unconstrained.

The TIP3P force field was used within LAMMPS to run an NPT MD trajectory of 520 waters for 1,000,000 steps of 0.25 fs each, with every 1,000th frame extracted to construct the pretraining dataset. A DFT dataset was generated from two separate AIMD trajectories performed in VASP with the SCAN functional using a similar procedure, resulting in 879 structures being extracted in total, 839 of which contain 65 waters and 40 of which contain 136 waters.[28][29] The SCAN meta-GGA functional was chosen due to its comparatively strong performance on liquid water when compared to other GGA or meta-GGA functionals.[30] Plane waves with energies up to 700 eV were used to construct the wavefunctions. The Brillouin zone was only sampled at the Γ-point owing to the large size, disordered structure, and insulating nature of the water molecules.

As with the QM9 case, three DFT-only training sets of different sizes (50, 100, and 500) were created by random selection to test the data efficiency improvements afforded by the T-PaiNN procedure. The remaining structures were used for hold-out testing. The accuracy of the various models on the water system was evaluated by comparing their energies, atomic forces and calculated pressures in addition to properties commonly extracted from molecular dynamics simulations, particularly predicted density, radial distribution functions, and water diffusivity, for which experimental values are well documented.

2.4 Code Development and Packages

All code was developed and written in python and MLIP construction and training was conducted using pytorch.[31] The SchNetPack python code was used for generating and executing the PaiNN MLIP architecture.[32] The Atomic Simulation Environment (ASE) package was used for running the MD simulation using the T-PaiNN derived MLIPs.[33]

## 3. Results and Discussion

To determine the viability of the T-PaiNN method, we benchmarked the transfer leaning capabilities on two systems: the QM9 benchmark small molecule dataset and simulating liquid water. The former demonstrates performance across a wide variety of molecular systems while the later demonstrates performance in complex inter-molecular systems. In all cases, the only input into the machine learned networks are the atomic positions, and the outputs are energies for the QM9 models and energies, forces, and stresses for the water models. From these, all other quantities are derived.

3.1 Small Molecule Energies

We examined the use of the T-PaiNN approach to non-interacting, i.e. gas phase, molecular systems through the QM9 dataset. The QM9 data was randomly partitioned into three sets:

pretraining, validation and test. A subset of 50,000 structures from the 134,000 in the QM9 database were randomly selected to form the pretraining set for calculation at the UFF level, as shown in Figure 2.a. This number was selected to ensure heterogeneity of molecules and systems without the need for calculating molecular dynamic trajectories. The validation set, used to detect overfitting and trigger early stopping of the training, comprised 20,000 molecules. The test set contained the remaining ~60,000 molecules.

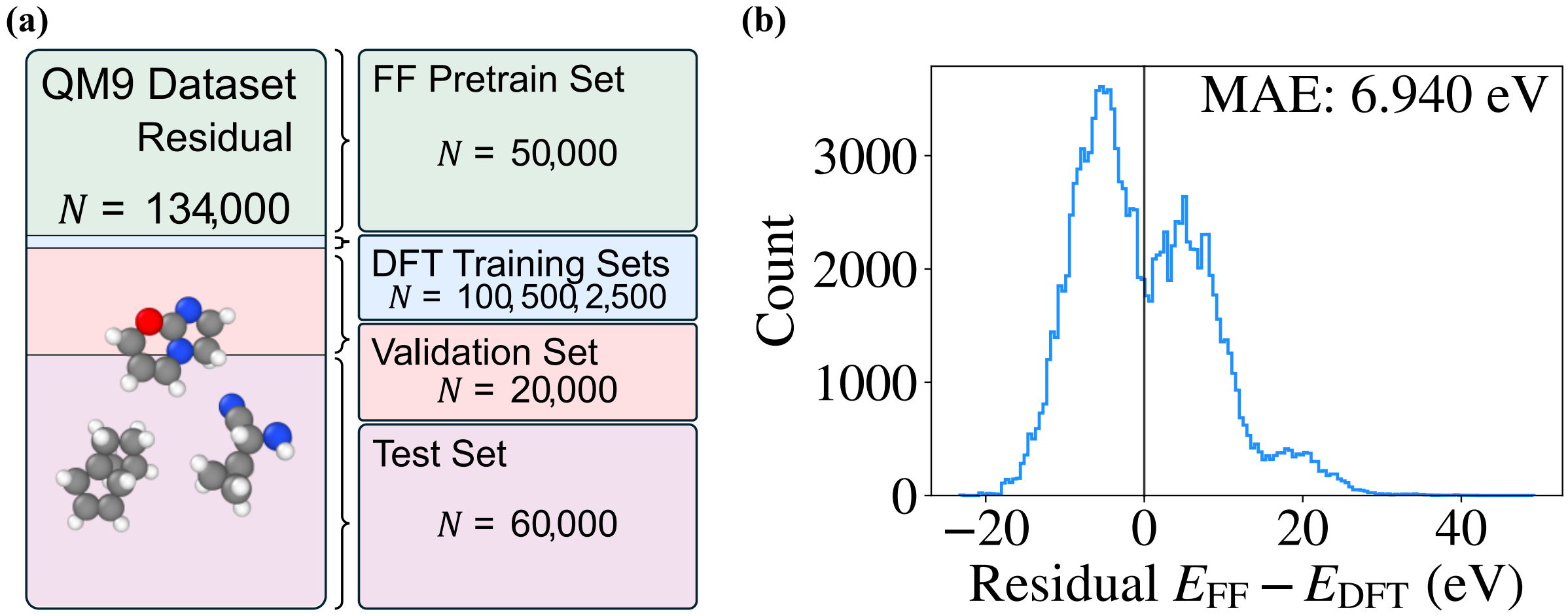

**Figure 2: (a)** A visualization of how the QM9 dataset was divided to form the sets used in this task, along with some randomly chosen example molecules from the dataset. **(b)** The energy residuals between the preexisting UFF force field with atomic energy offsets added and the corresponding DFT energies for all molecules in the QM9 dataset.

*3.1.1 Generation of Pretraining Set and Alignment of Energies*

To align the classical force field and DFT potential energy surfaces, elemental atomic energy offsets were calculated via linear regression between molecule element counts and their UFF to DFT energy residuals. This was performed using only molecules in the DFT training sets, of sizes 100, 500 or 2,500 depending on the DFT training set used for autotuning. The residual distribution of the UFF energies aligned to 500 DFT examples is shown in Figure 2.b. The mean absolute error of 6.9 eV is a 57% improvement from the atomic-offset-only MAE of 12.2 eV (i.e. what is expected just based on the number of each element and their energies). This improvement demonstrates that the classical force field partially captures the physics of the molecules; however, the 6.9 eV error is insufficient for examining detailed chemistry. The error distribution is bimodal; the negative mode peaks near -5.16 eV and predominantly comprises pure hydrocarbons while the positive mode peaks near +4.85 eV and predominantly contains N and O-rich molecules. These large and systematic errors indicate that the force field is doing a poor job characterizing the detailed chemistry of these molecules, despite getting some of the behavior correct. Nevertheless, these energies provide basic information which is sufficient to serve as the initial T-PaiNN model training set.

*3.1.2 Model Accuracy and Training Speed of T-PaiNN on Molecular Systems*

To test the improvement in data efficiency afforded by the T-PaiNN procedure, three small PaiNN models (36.3k parameters) were generated and trained on the 100, 500, and 2,500 molecule datasets to serve as DFT-only models. These models are referred to by the number of exemplar DFT molecules included. An identical PaiNN model (36.3k parameters) was trained on the $N$ = 50,000 UFF dataset and served as the standard initial state for T-PaiNN models, which was autotuned on each of the three DFT datasets. This procedure resulted in a total of seven PaiNN models: three DFT-only models trained only on varying amounts DFT data, one model based on the UFF only dataset, and three T-PaiNN models trained on the 50,000 UFF dataset and then autotuned on the three DFT training sets. As these models trained, they were evaluated on a completely held-out test set of 60,000 molecules from the QM9 dataset, the resulting training loss curves for both the DFT-only PaiNN models and the T-PaiNN models are shown in Figure 3.

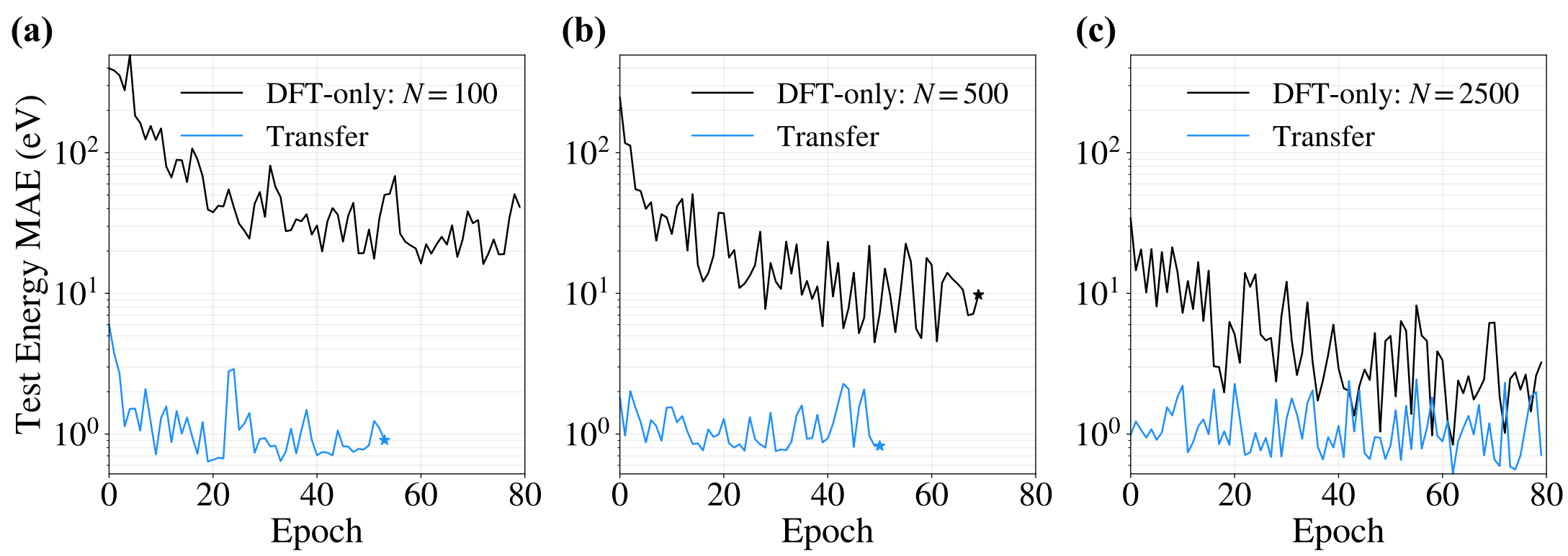

**Figure 3**: QM9 test-set performance for the DFT-only (black) and T-PaiNN (blue) GNN MLIPs over the course of their training. The curves vary in the number of epochs due to validation set triggered early-stopping during training, which is indicated by a star at the end of the curve.

For all training set sizes, the T-PaiNN models outperform the DFT-only models in terms of both global minimum MAE on the test set and the number of DFT training epochs required to achieve optimal performance. The DFT-only PaiNN models achieved minimum test set MAE values of 16.2, 4.5, and 0.8 eV from the 100, 500, and 2,500 DFT training sets respectively, while T-PaiNN models achieved minimum MAE values of 0.6, 0.8, and 0.5 eV when trained on the same DFT training sets. These correspond to 25.3×, 5.9×, and 1.6× reductions in MAE for the same amount of expensive DFT calculations.

In addition to this dramatically improved performance, the T-PaiNN models reached their accuracy-plateau more quickly than the DFT-only models. Model training speed-up is quantified by finding the first epoch where test MAE is within 10% of the minimum value. The T-PaiNN method showed a 5.7×, 3.5×, and 2.0× speedup on the 100, 500, and 2,500 sized training sets, achieving convergence in 9, 14, and 31 epochs respectively compared to 51, 49, and 62 for the DFT-only models. Thus, the T-PaiNN method not only increases accuracy, but achieves convergence faster than the simple DFT only method.

### *3.1.3 Model Size Effects of T-PaiNN*

The T-PaiNN procedure performance was evaluated as a function of model size with medium and large sized PaiNN models (197k and 534k parameters, respectively as shown in Table 1). This test examined just the $N = 500$ DFT dataset for both T-PaiNN and the DFT only PaiNN models. The training accuracy results are shown in Figure 4 and are similar to those found with the smaller models, but with lower error for both the T-PaiNN and DFT only systems. The DFT-only model reached a minimum test set MAE of 1.45 eV on epoch 59, while the T-PaiNN model reached a minimum of 0.23 eV on epoch 98, a 6.3× improvement. It is crucial to note that with the 0.23 eV error mark, T-PaiNN is approaching expected B3LYP accuracy to the true chemical energies; conversely, the 1.45 eV MAE of the DFT only PaiNN model renders it useless for examining chemistry.

Across the 100 training epochs, only three individual T-PaiNN epochs (at the vary beginning) had test set MAEs worse than the best DFT-only model at its end. Additionally, from epochs 50 to 100 (chosen to capture the accuracy plateaus for both models), the T-PaiNN model demonstrated much greater stability than the DFT-only model, with a standard deviation of 0.4 eV in MAE compared to 3.2 eV. These results indicate both improved performance and stability of the T-PaiNN models compared to DFT-only training. All results for the molecular systems are aggregated in Table 2. Overall, the T-PaiNN method greatly increases the accuracy of GNN-MLIPs at minimal extra cost of data collection, while simultaneously smoothing epoch-to-epoch training variance and decreasing time to DFT training convergence in molecular systems.

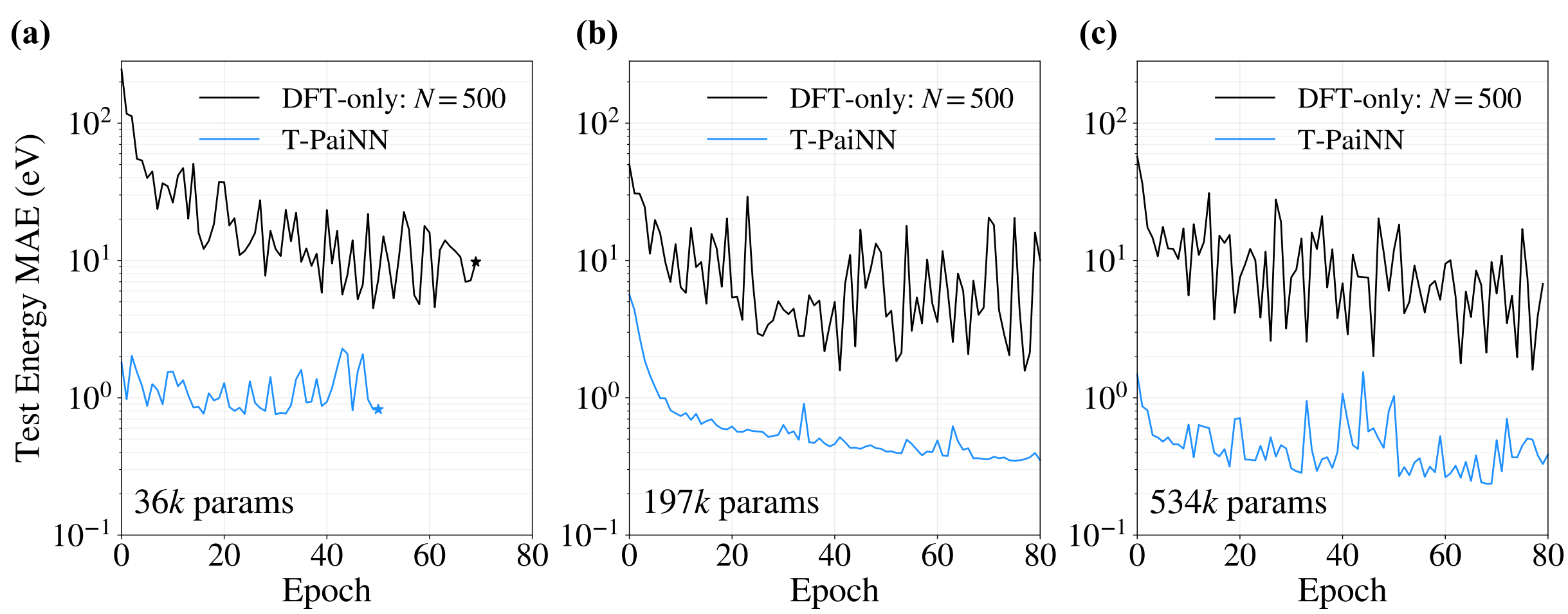

**Figure 4**: QM9 test-set performance for DFT-only (black) and T-PaiNN (blue) PaiNN models of **(a)** small (36k parameters), **(b)** medium (197k parameters), and **(c)** large (534k parameters), as they were trained on the $N = 500$ dataset. The curves vary in the number of epochs due to early stopping being used during training based on the validation data, which is indicated by a star at the end of the curve.

**Table 2**: Summary of best performance achieved on the QM9 test set for each model.

| DFT Training Set Size | Model Parameters | Minimum MAE (eV) | | | Epochs to Convergence | | |
|---|---|---|---|---|---|---|---|
| | | DFT-only | T-PaiNN | Improvement | DFT-only | T-PaiNN | Improvement |
| 100 | 36,300 | 16.2 | 0.6 | 25.3× | 51 | 9 | 5.7× |
| 500 | 36,300 | 4.49 | 0.76 | 5.9× | 49 | 14 | 3.5× |
| 2,500 | 36,300 | 0.84 | 0.52 | 1.6× | 62 | 31 | 2.0× |
| 500 | 197,000 | 1.13 | 0.32 | 3.5× | 92 | 67 | 1.6× |
| 500 | 534,000 | 1.60 | 0.23 | 6.3× | 77 | 63 | 1.2× |

## 3.2 Liquid Water Behavior

A major hurdle in examining aqueous chemistry is the representation of solvated systems. To do so has conventionally required trade-off between low-cost, lower-accuracy implicit solvent methods and higher-cost, higher-accuracy explicit solvation. MLIPs are highly promising for explicitly represent solvating systems while retaining quantum level accuracy. However, gathering sufficient quantum data across vast configuration space associated with liquids with which to construct the MLIP is daunting. We aim to examine the ability of T-PaiNN to overcome this challenge using water as an exemplar system. For this section, only the large PaiNN architecture was used, see Table 1 for hyperparameters.

Unlike the QM9 task, which required only the prediction of molecular energies, this task examines the generation of a model that can perform molecular dynamics runs of water, and therefore high accuracy energies, forces, and simulation cell stresses are all required. Forces and stresses are both properties derived from differentiating the potential energy function of the system with respect to atomic coordinates or cell lattice parameters, respectively, both of which can easily be obtained from a GNN-MLIP by autodifferentiation.

### *3.2.1 Alignment of Classical and Quantum Values for Liquid Water System*

The TIP3P force field behavior correlates well with the SCAN DFT calculations, as shown by the correlation between the predicted energy, force magnitude, and force direction by the two methods, shown in Figure 5. Virial stress values (aggregated into pressures) show considerable correlation ($R^2 = 0.64$), but a large MAE of 2376 bar. This disagreement arises from the known bias of TIP3P to generally under-predict the density of liquid water and the tendency of SCAN-DFT to over-predict the density. However, the underlying correlation is sufficient to translate the classical energies, forces and stresses into the DFT energy scale.

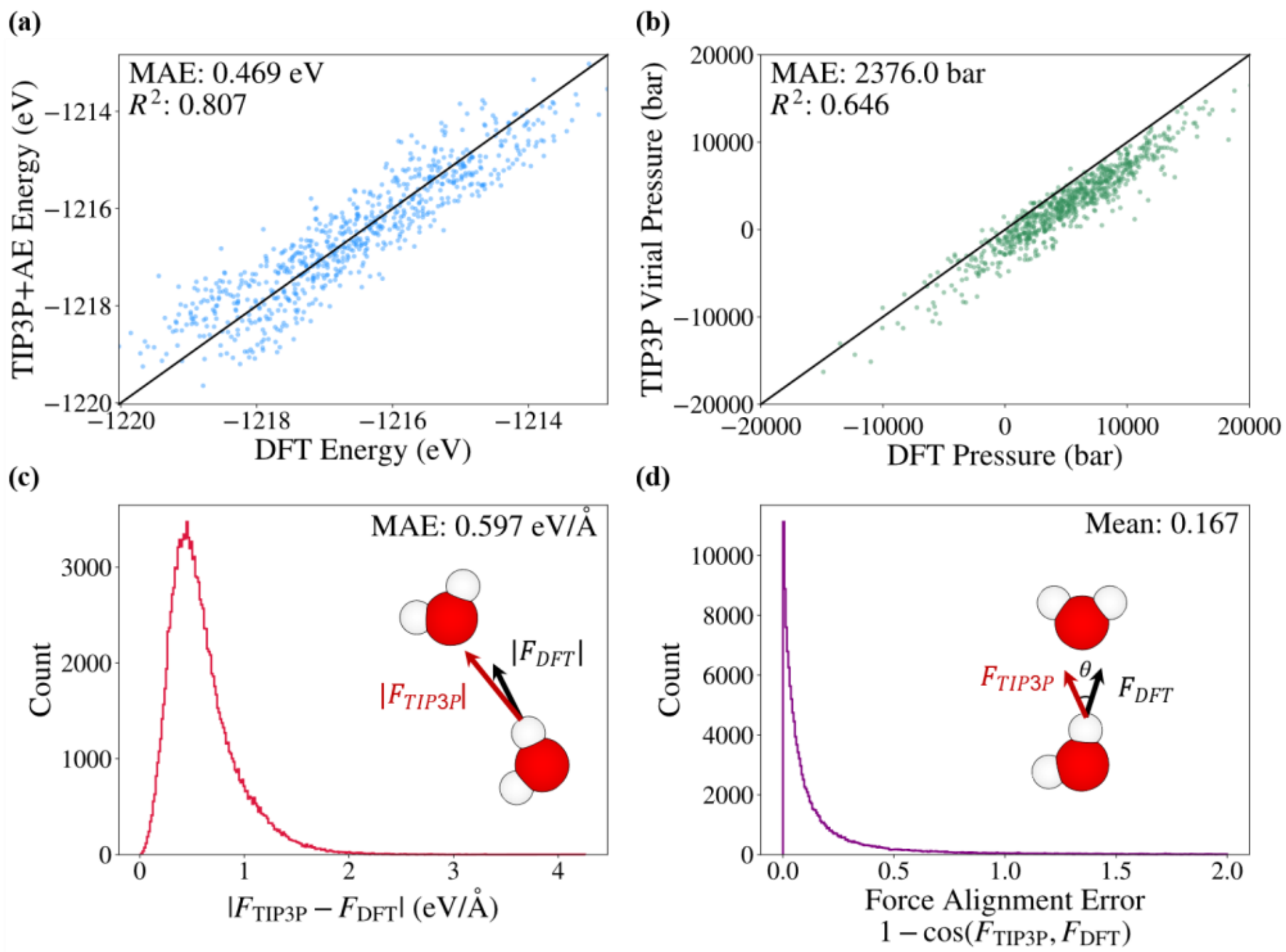

**Figure 5:** Comparison of TIP3P energies, forces, and stresses (aggregated into pressures) to those of DFT on the same structures. **(a)** Shows the parity plot between atomic energy adjusted TIP3P energies and DFT energies, **(b)** shows the parity of the pressures of these structures, **(c)** shows a histogram of force magnitude errors, and **(d)** shows a histogram of force vector alignment errors.

3.2.2 Liquid Water T-PaiNN Performance with Varying Amounts of DFT Data

The training accuracy of the different models are shown in Figure 6. Similarly to the QM9 task, the T-PaiNN models demonstrate improved performance across the board. In terms of energetics (Figure 6.a-c) the DFT-only models showed minimum test set MAE values of 0.49, 0.46, and 0.39 eV from the $N = 50$, 100, 500 datasets respectively, while the T-PaiNN models had minimum test set MAEs of 0.33, 0.30 and 0.27 eV. From epochs 100 to 200 (chosen to represent the accuracy plateau region), the T-PaiNN models demonstrated improved stability, with standard deviations in test set MAE of 0.32, 0.36, and 0.27 eV, in comparison to 1.89, 1.95, and 0.95 for the DFT-only models. Thus, in addition to the improvements in MAE, the T-PaiNN models appear to train within a smoother and more stable region of parameter space. This is particularly encouraging for the extrapolation ability of T-PaiNN trained models, and the stability of long molecular dynamics runs. We hypothesize that the significant, but lower factional, improvement of T-PaiNN on the water system compared to the QM9 system is that water is the only molecule which needed to be learned by the MLIP.

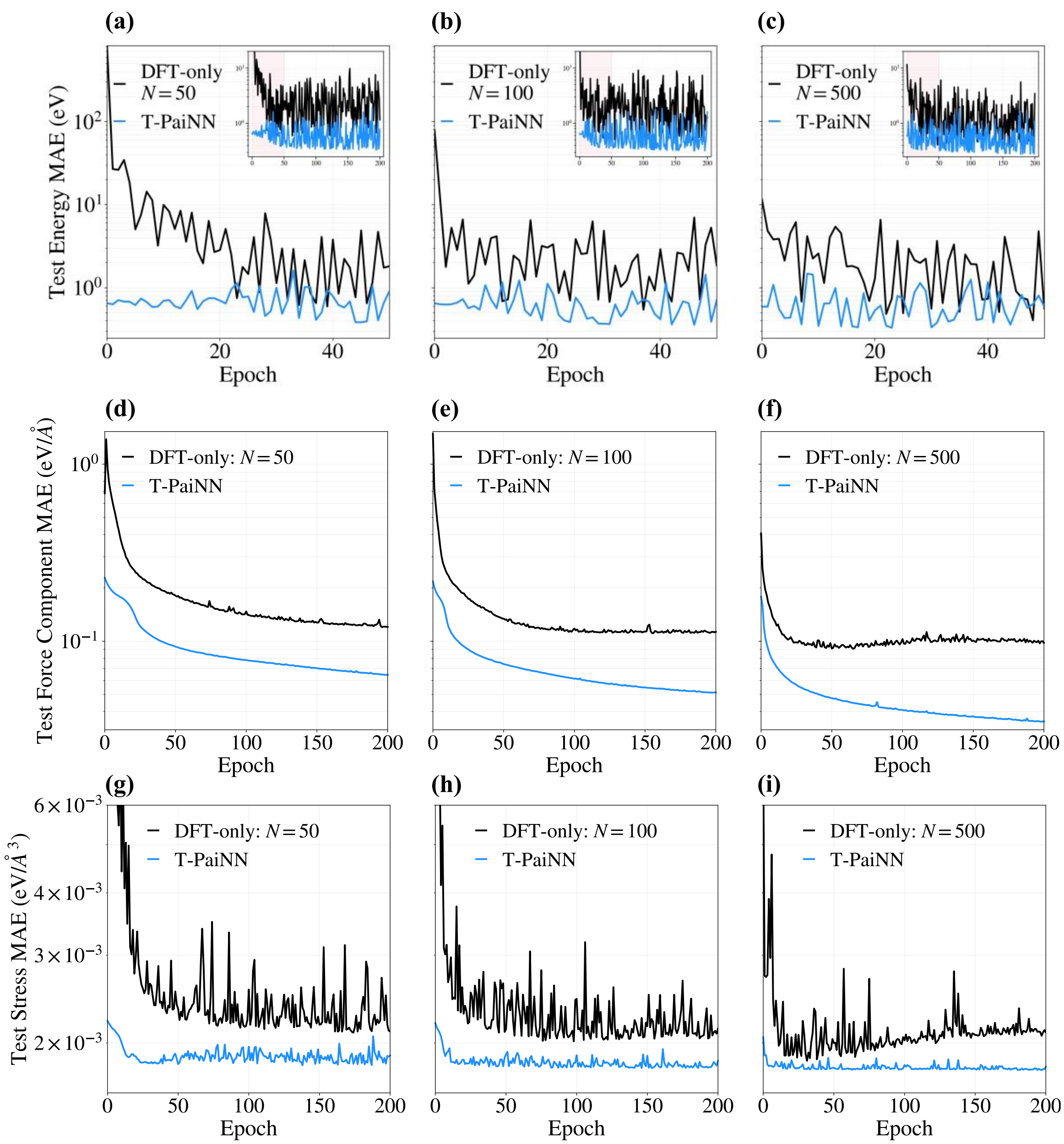

**Figure 6**: DFT water test set performance for the DFT-only (black) and T-PaiNN (blue) models during their training on three different sized DFT training sets, $N$ = 50 **(a,d,h)**, 100 **(b,e,i)**, and 500 **(c,f,j)**. The first row **(a,b,c)** shows energy MAE, the second **(d,e,f)** shows force vector component MAE, and the third **(h,i,j)** shows stress component MAE. For the energy subplots, the first 50 epochs are shown in the main plots, highlighting the early training improvements, with all 200 epochs shown in the inset.

T-PaiNN models demonstrate even stronger improvements in terms of calculated force as shown by the MAEs in Figure 6.d-f. The three DFT-only models reached minimum MAEs of 0.12, 0.11, and 0.09 eV / Å when trained on the $N$ = 50, 100, 500 datasets respectively, while the corresponding T-PaiNN models were approximately half that, reaching minimum MAEs of 0.06, 0.05, and 0.04 eV / Å. For all T-PaiNN models, the force MAEs continued to improve until the end of the 200 epoch training, with the minima found at epochs 199, 197 and 198 for the $N = 50$, 100, and 500 models respectively. This was also true for the $N = 50$ DFT-only model, but for the other two the force MAEs stagnated or began to increase for the $N = 100$ and 500 DFT-only models respectively, indicating overfitting which the corresponding T-PaiNN models did not experience.

The stress MAE training curves shown in Figure 6.h-j show similar behavior to the force MAE curves. The three DFT-only models demonstrated MAE minima of 2.1, 2.0, and 1.8 $\times 10^{-3}$ eV / Å$^3$ for the $N = 50$, 100, and 500 models, with the corresponding T-PaiNN models demonstrating minima of 1.8, 1.7, and 1.7 $\times 10^{-3}$ eV / Å$^3$. In this case, all three DFT-only models either stagnated or began to degrade in stress performance over the course of the 200 epoch training. The three T-PaiNN models conversely do not demonstrate any degradation of performance, with the $N = 50$ model stagnating, but the $N = 100$ and 500 T-PaiNN models demonstrate continued improvement, with their minima occurring at epochs 186 and 197 out of 200.

*3.2.2 Molecular Dynamics Simulation of Water using T-PaiNN and DFT-only PaiNN MLIP*

While the liquid water results of Section 3.2 suggest that the T-PaiNN procedure improves the prediction of DFT water energies, forces, and stresses, a more important question is whether these gains translate into enhanced physical predictions in molecular dynamics simulation. Molecular dynamics simulation of liquid water therefore provides a more relevant test of the T-PaiNN methodology than merely hold-out predicted energy and force error alone. The performance of three ML force field models, a DFT-only PaiNN, a TIP3P-only PaiNN (pretrained), and T-PaiNN, are examined for accuracy across three experimental observables: water density, the water radial distribution functions, and water self-diffusion. The results are shown in Figure 7.

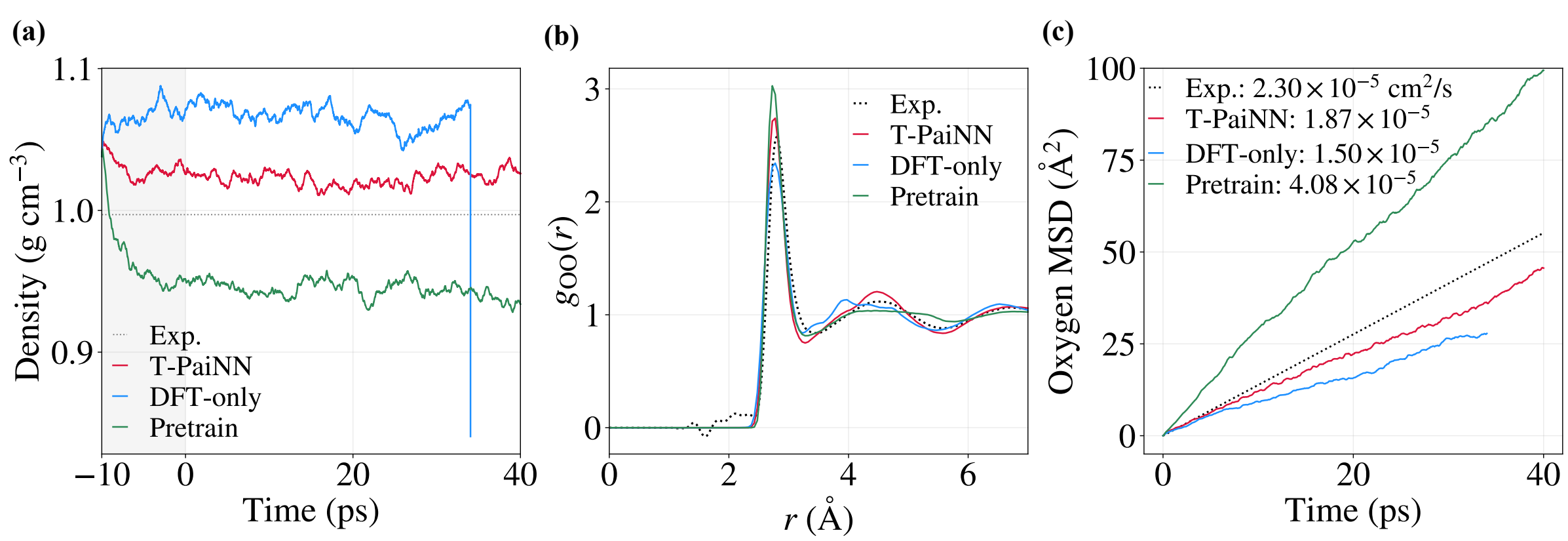

**Figure 7**: The performance of the DFT-only PaiNN, TIP3P-only PaiNN, and T-PaiNN models in 50ps NPT MD simulations of a box of 520 waters. **(a)** The density of the water box for the three water models, **(b)** their radial distribution functions, **(c)** the mean square displacement of the oxygen atoms over the simulations.

Across all three experimental observables, the T-PaiNN MLIP properties match closer to experiment than the other two PaiNN models, as tabulated in Table 3. The PaiNN model derived from the classical force field pre-training underestimates the density while overestimating the hydration shell structuredness (higher peak around 3Å in the RDF of Figure 7.b), and self-diffusivity of water. The DFT-only trained PaiNN model overestimates the water destiny and underestimates the structural rigidity and self-diffusion. These are known limitations of the TIP3P and SCAN-DFT water PESs. T-PaiNN is closer to experiment than the other two across all metrics aside from four (see Table 3): the intensities of the RDF peaks as calculated by DFT only PaiNN, which performs worse than T-PaiNN in predicting peak location, and the first peak location of TIP3P. Given the importance of peak location, we posit that a <10% error in peak intensity compared to experiment and DFT is acceptable, given the very good agreement with peak location. Beyond being the model which best reproduces experiment, the T-PaiNN model is more similar to the DFT-only model than the pretrained PaiNN model. Thus, we conclude that the transfer learning processes embeds the model with the key energetic information from DFT and that the improved performance compared to DFT-only arises from being exposed to a larger number of configurations.

**Table 3**: Comparison of DFT-only and T-PaiNN models performance on the liquid water task.

| Metric | | | | DFT-only | TIP3P-only | T-PaiNN | Experiment | Improvement |
|---|---|---|---|---|---|---|---|---|
| Intrinsic† | Energy (eV MAE) | | | 0.36 | 1.08 | **0.26** | - | 1.40× |
| | Force (eV/Å MAE) | | | 0.09 | 0.23 | **0.03** | - | 2.80× |
| | Stress (eV/$Å^3$ MAE) | | | 0.0018 | 0.0023 | **0.0017** | - | 1.05× |
| Derived‡ | Density ($g/cm^3$) | | | 1.07 | 0.94 | **1.02** | 0.99 | - |
| | Self-Diffusion ($\times 10^{-5}$ $cm^2/s$) | | | 1.50 | 4.08 | **1.87** | 2.30 | - |
| | RDF Extrema | 1st Peak | Position | 2.78 | 2.78 | 2.78 | 2.80 | - |
| | | | Intensity | 2.34 | 2.95 | **2.73** | 2.58 | - |
| | | 1st Trough | Position | 3.28 | **3.33** | 3.28 | 3.47 | - |
| | | | Intensity | **0.84** | 0.82 | 0.75 | 0.84 | - |
| | | 2nd Peak | Position | 3.98 | 4.38 | **4.48** | 4.51 | - |
| | | | Intensity | **1.13** | 1.03 | 1.20 | 1.12 | - |
| | | 2nd Trough | Position | 5.48 | 4.48 | **5.58** | 5.58 | - |
| | | | Intensity | **0.86** | 1.04 | 0.84 | 0.88 | - |
| | | 3rd Peak | Position | 6.58 | 6.78 | **6.73** | 6.69 | - |
| | | | Intensity | 1.09 | 1.03 | **1.07** | 1.06 | - |
| | | | Wins | 3 | 1 | **10** | | |

† Comparison to SCAN-DFT evaluated held-out test set.

‡ Derived from 50ps NPT molecular dynamics simulation of 520 waters.

# 4. Conclusions

Herein, we have developed and demonstrated a novel Transfer-PaiNN method for training interatomic graph neural network potentials. By pretraining network with low-cost, low-accuracy classical force filed data, and subsequently autotuning the resulting model with DFT data, the resulting autotuned T-PaiNN model significantly and persistently outperforms models trained on DFT data alone. The improved performance holds across molecular and periodic boundary condition systems and systems with intra- *and* inter-molecular forces. We find performance improvements of 2-10×, and that this method brings the error closer to the DFT expected accuracy. We hypothesize that this performance improvement arises from the model learning general interatomic interactions from a much larger range of systems. Thus, the model is less likely to rely on extrapolation than DFT only models. Overall, these results show that T-PaiNN is a practical methodology for significantly improving graphical neural networks for atomic calculations with minimal additional computational cost.

# 5. Funding Sources

The Authors gratefully acknowledge several funding sources which supported the authors contributing to this cross-project collaborative effort. VP acknowledges support from the Science and Technologies for Phosphorus Sustainability (STEPS) Center, a National Science Foundation Science and Technology Center (CBET-2019435). VB acknowledges support from U.S. Department of Energy, Office of Science, Office of *Basic Energy Sciences*, under Award Number DE-SC0024194. CM and DR acknowledge support from the U.S. National Science Foundation CBET Catalysis under award 2450869. SR acknowledges support from the DOE Office of Science, Office of Basic Energy Sciences (BES), Materials Sciences and Engineering Division under Award DE-SC0024724. The content is solely the authors' responsibility and does not necessarily represent the official views of the National Institutes of Health, US Department of Energy, or the NSF. In addition, we acknowledge support from Research Computing at Arizona State University for providing high-performance supercomputing services.